\documentclass[12pt]{article}

\usepackage{times}

\usepackage[colorlinks=true,bookmarks=false,citecolor=blue,urlcolor=blue]{hyperref} 
\usepackage{amsmath,amssymb, mathrsfs}
\usepackage{graphicx}
\usepackage{float}
\usepackage{xcolor}
\usepackage{soul}

\usepackage{capt-of}

\usepackage[
    backend=biber,
    style=nature,
  ]{biblatex}

\newenvironment{sciabstract}{%
\begin{quote} \bf}
{\end{quote}}

\title{Enhanced sensitivity via non-Hermitian topology}

\author{Midya Parto$^{1,\ast}$, Christian Leefmans$^{2,\ast}$, James Williams$^1$, Alireza Marandi$^{1,2,\dagger}$\\
\\
\normalsize{$^1$Department of Electrical Engineering, California Institute of Technology, Pasadena, CA 91125, USA.}\\
\normalsize{$^2$Department of Applied Physics, California Institute of Technology, Pasadena, CA 91125, USA.}\\
\\
\normalsize{$^\ast$These authors contributed equally}\\
\normalsize{$^\dagger$marandi@caltech.edu}
}

\date{}

\begin{document} 


\baselineskip24pt


\maketitle 

\begin{sciabstract}
Sensors are indispensable tools of modern life that are ubiquitously used in diverse settings ranging from smartphones and autonomous vehicles to the healthcare industry and space technology \cite{riemensberger_massively_2020,kim_wearable_2019,mangold_perseverance_2021}. By interfacing multiple sensors that collectively interact with the signal to be measured, one can go beyond the signal-to-noise ratios (SNR) than those attainable by the individual constituting elements. Such distributed sensing techniques have also been implemented in the quantum regime, where a linear increase in the SNR has been achieved via using entangled states \cite{guo_distributed_2020}. Along similar lines, coupled non-Hermitian systems \cite{el-ganainy_non-hermitian_2018,parto_non-hermitian_2021} have provided yet additional degrees of freedom to obtain better sensors via higher-order exceptional points \cite{hodaei_enhanced_2017,chen_exceptional_2017}. Quite recently, a new class of non-Hermitian systems, known as non-Hermitian topological sensors (NTOS) has been theoretically proposed \cite{budich_non-hermitian_2020,koch_quantum_2022}. Remarkably, the synergistic interplay between non-Hermiticity and topology is expected to bestow such sensors with an enhanced sensitivity that grows exponentially with the size of the sensor network. Here, we experimentally demonstrate NTOS using a network of photonic time-multiplexed resonators in the synthetic dimension represented by optical pulses. By judiciously programming the delay lines in such a network, we realize the archetypal Hatano-Nelson model \cite{hatano_localization_1996} for our non-Hermitian topological sensing scheme. Our experimentally measured sensitivities for different lattice sizes confirm the characteristic exponential enhancement of NTOS. We show that this peculiar response arises due to the combined synergy between non-Hermiticity and topology, something that is absent in Hermitian topological lattices. Our demonstration of NTOS paves the way for realizing sensors with unprecedented sensitivities.
\end{sciabstract}

The ability to accurately and reliably measure physical quantities is at the heart of modern sensors with applications ranging from molecular sensing in chemistry \cite{xue_solid-state_2020} and biology \cite{altug_advances_2022} to light detection and ranging (LiDAR) \cite{zhang_large-scale_2022} and observing gravitational waves \cite{aasi_enhanced_2013}. Significant efforts have been made towards enhancing the response of individual sensing elements, for instance by using high-quality resonators \cite{zhu_-chip_2010} or exploiting quantum effects \cite{degen_quantum_2017}. A different, more generic route to achieving higher sensitivities is to employ a multitude of modes that collectively contribute to a coherent signal that encapsulates information about the quantity to be measured. This has led to distributed classical and quantum sensing networks which allow for an enhancement of $\sqrt{N}$ and $N$ \cite{liu_distributed_2021} in the sensitivity figure, as compared to a single sensing element, respectively. 

An alternative path to achieve higher sensitivities is by employing concepts from non-Hermitian physics \cite{el-ganainy_non-hermitian_2018,parto_non-hermitian_2021,ashida_non-hermitian_2020}. For instance, the eigenvalues associated with a non-Hermitian system can respond to perturbations in a remarkably stronger manner compared to its Hermitian counterparts. This realization is the foundation of a class of sensors that operate in the vicinity of non-Hermitian degeneracies known as exceptional points (EPs), where the corresponding response scales as the $N$-th root of the perturbation with the order of the EP \cite{hodaei_enhanced_2017,chen_exceptional_2017,kononchuk_exceptional-point-based_2022}. In addition, the introduction of non-Hermiticity to topologically non-trivial lattices is known to result in an eigenspace that behaves very differently from that associated with Hermitian topological systems \cite{,gong_topological_2018,bergholtz_exceptional_2021}. Recent studies have observed the manifestation of this distinct behavior in the form of a new type of bulk-boundary correspondence and the non-Hermitian skin effect \cite{kunst_biorthogonal_2018,yao_edge_2018,xiao_non-hermitian_2020,helbig_generalized_2020,weidemann_topological_2020}.

Quite recently, a new class of sensors based on the synergy between non-Hermiticity and topology has been proposed \cite{budich_non-hermitian_2020,koch_quantum_2022}. Dubbed as non-Hermitian topological sensors (NTOS), such devices can exhibit a sensitivity that grows exponentially with respect to the number of lattice sites. Remarkably, unlike typical non-Hermitian sensing schemes, this boosted response does not require fine tuning of the system parameters. Despite intense activities in the field of non-Hermitian topology, an experimental observation of this enhanced sensitivity has so far remained elusive. Here, we experimentally demonstrate this peculiar behavior in a network of photonic time-multiplexed resonators. By using different number of optical pulses, we realize non-Hermitian topological lattices with number of lattice sites up to $N=23$. Based on measurement results from these structures, we experimentally demonstrate the characteristic exponential growth of the sensitivity associated with NTOS. It will be shown that this extraordinary response arises exclusively due to the cooperative interplay between non-Hermiticity and topology, something that is absent in other Hermitian topological settings.


\begin{figure}
    \centering
    \includegraphics[width=1\textwidth]{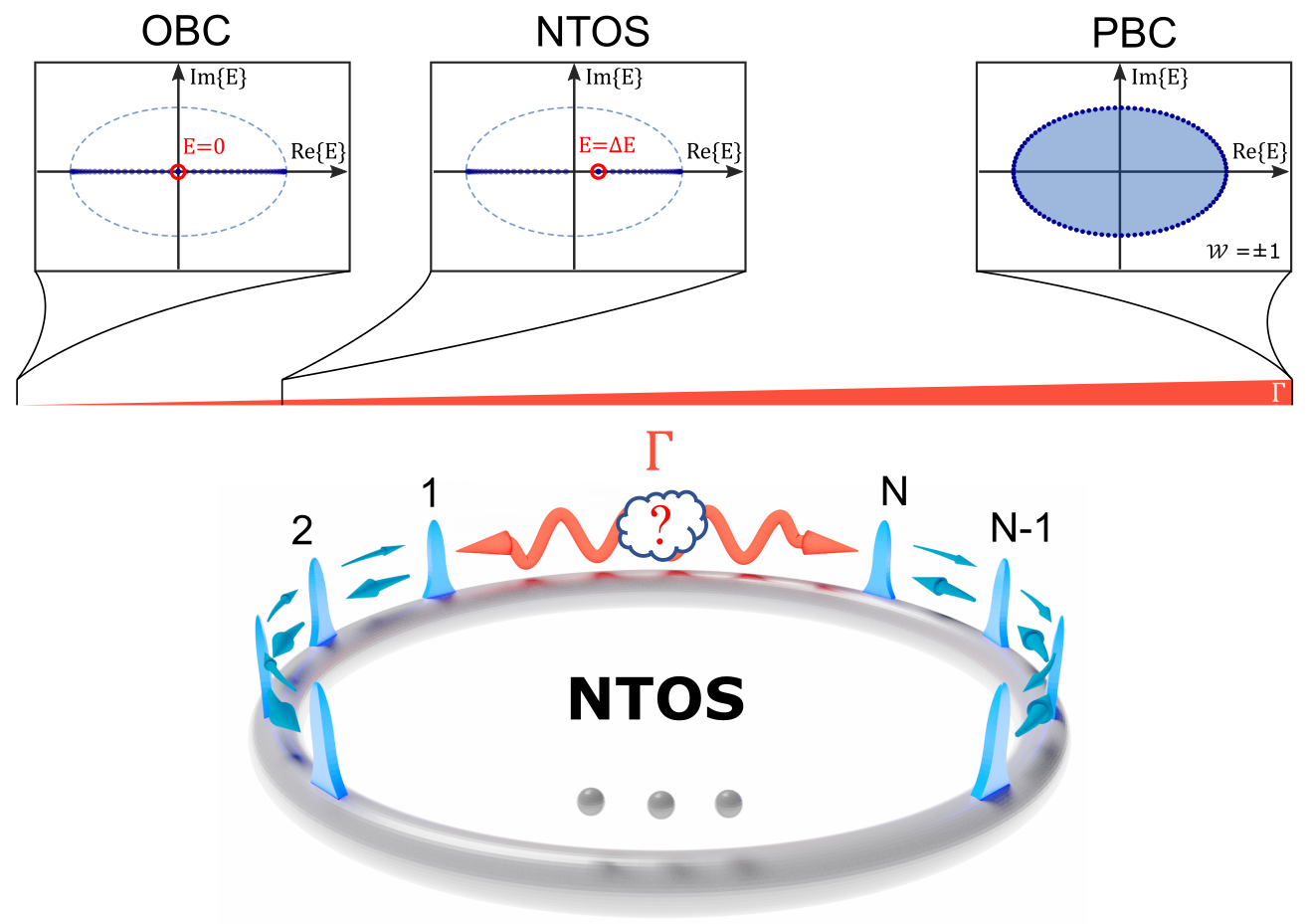}
    \caption{\textbf{Non-Hermitian topological sensors (NTOS).} Schematic diagram of the NTOS demonstrated here based on the Hatano-Nelson model which features nonreciprocal couplings between the adjacent elements of the array. Depending on the boundary conditions, this lattice exhibits different eigenvalue spectra, as shown in the top part of the figure. This can be represented by the strength $\Gamma$ of the coupling between the first and last resonators in the system. When $\Gamma$ is equal to the other couplings in the array (the rightmost part of the scale), the structure follows periodic boundary conditions (PBC), where the eigenvalues form an ellipse around the origin in the complex plane. In this case, a nonzero winding number $\mathcal{W}$ can be defined. On the other hand, when $\Gamma=0$, i.e. under open boundary conditions (OBC), all the eigenvalues reside on the real axis, with one eigenvalue exactly equal to zero $E=0$ (for odd values of $N$). This eigenvalue tends to shift from its original value by $\Delta E$ which is proportional to the strength of the boundary coupling $\Gamma$, as long as the coupling is sufficiently small. This mechanism can be effectively harnessed for sensing any perturbation that modifies $\Gamma$.}
    \label{NTOS}
\end{figure}

For our realization of NTOS, we consider the Hatano-Nelson \cite{hatano_localization_1996} model as described by the Hamiltonian: 

\begin{equation}
    \hat{H}_{HN}=\sum_{n}t_{\rm R} \hat{a}_{n+1}^\dagger\hat{a}_n+t_{\rm L}\hat{a}_{n}^\dagger\hat{a}_{n+1},
\label{eq:HN}
\end{equation}
where $\hat{a}_n^{(\dagger)}$ is the annihilation (creation) operator associated with site $n$ while $t_{\rm R}$, $t_{\rm L}$ represent the nonreciprocal right and left nearest-neighbor couplings within the lattice. When truncated corresponding to a finite lattice, the Hamiltonian of Eq. \ref{eq:HN} can exhibit a multitude of spectral behaviors, depending on the associated boundary conditions (Fig. \ref{NTOS}). In particular, when the lattice is arranged in a uniform fashion with periodic boundary conditions (PBC), the set of eigenvalues form a closed loop in the complex plane with a nonzero winding around the origin (Fig. \ref{NTOS}), associated with uniformly distributed bulk eigenstates across the array. We would like to emphasize that here, since the coupling mechanism between lattice elements are dissipative \cite{leefmans_topological_2022}, the real part of the system eigenvalues represent dissipation while the imaginary part corresponds to phase/frequency shift. On the other hand, when the structure is terminated with open boundary conditions (OBC), the resulting spectrum is entirely real (Fig. \ref{NTOS}). This corresponds to the case where all the eigenstate become localized on one edge of the system, known as the non-Hermitian skin effect \cite{weidemann_topological_2022}. Furthermore, under such OBC conditions, provided that the number of elements in the lattice is odd $N=2k+1$, the Hamiltonian $\hat{H}_{HN}$ always possesses an eigenstate $\left|\psi_{0}\right>_R$ with an eigenvalue equal to zero. 

\begin{figure}
    \centering
    \includegraphics[width=1\textwidth]{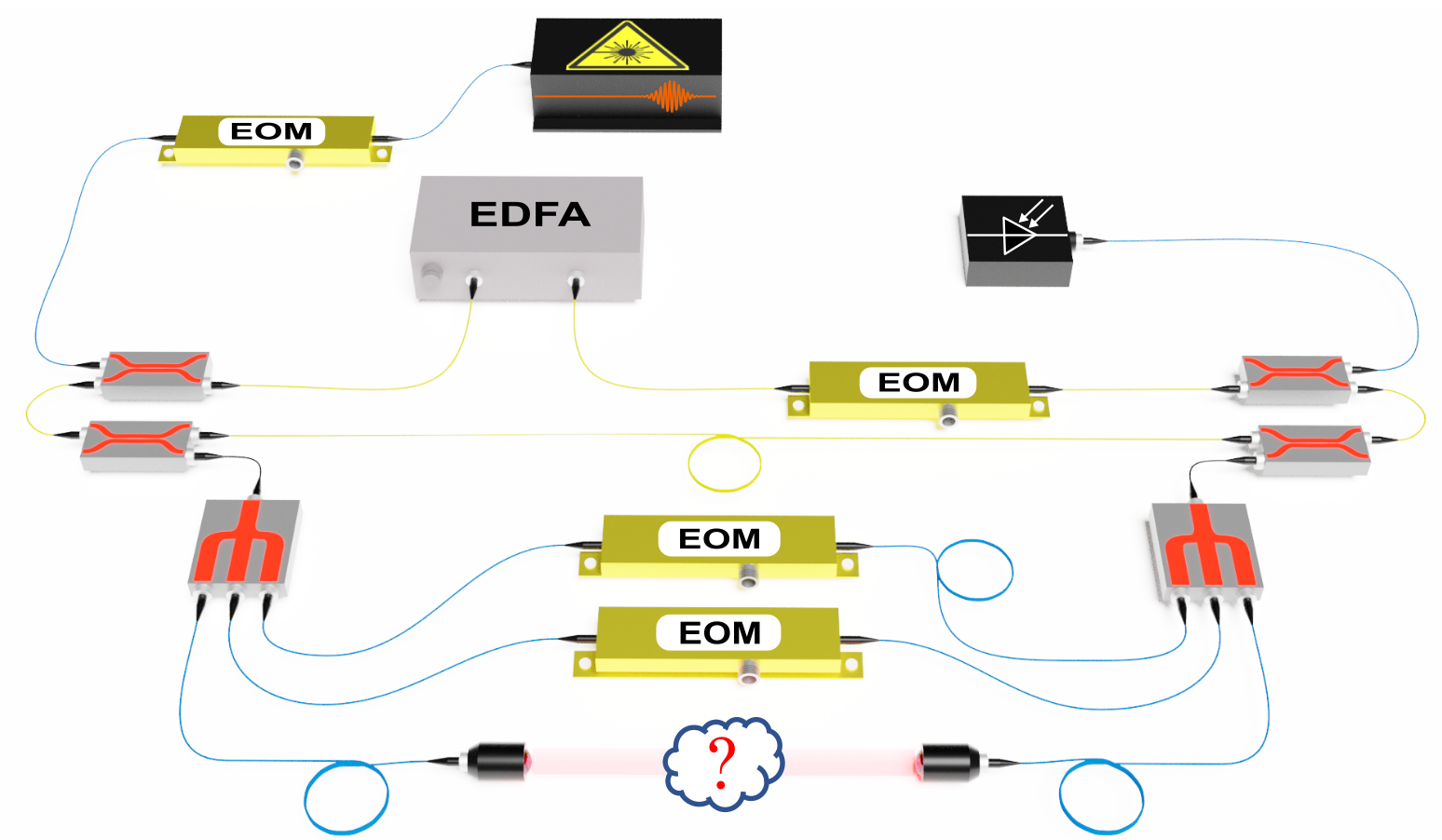}
    \caption{\textbf{Schematic of the network of time-multiplexed resonators used to demonstrate NTOS.} Synthetic resonators are defined by femtosecond pulses emitted by a mode-locked laser with a repetition rate of $T_R$ passing through an electro-optic modulator (EOM) before injection into the optical fiber-based cavity (yellow fibers). An Erbium-doped fiber amplifier (EDFA) is used in the main cavity to compensate for the losses and increase the number of measurement roundtrips. Two delay lines with smaller and larger lengths than the main cavity (corresponding to delays of $-T_R$ and $+T_R$, respectively) are utilized to provide nonreciprocal couplings between the nearest-neighbor resonators, necessary to implement the non-Hermitian topological model of Eq. \ref{eq:HN}. In addition, a third delay line with a length that corresponds to an optical delay of  $+(N-1)T_R$ associated with the perturbation $\Delta \hat{H}$ is also included. The strength of such a perturbation, i.e. $\Gamma$, can be accurately adjusted via a controlled misalignment in the free space section depicted in the figure.}
    \label{Schematic}
\end{figure}

To experimentally demonstrate NTOS, we use a time-multiplexed photonic resonator network depicted schematically in Fig. \ref{Schematic}. The network consists of a main fiber loop which supports $N$ resonant pulses separated by a repetition period, $T_{\rm R}$. Here, each individual pulse represents a single resonator associated with the annihilation (creation) operators $\hat{a}_j^{(\dagger)}$ in Eq.~\ref{eq:HN}. To realize the non-reciprocal couplings $t_{\rm R}$ and $t_{\rm L}$, we use delay lines to dissipatively couple nearest-neighbor pulses. Each delay line is equipped with intensity modulators that control the strengths of such couplings (see Fig. \ref{Schematic}). To induce the perturbation signal, we consider a change in the lattice of the form $\Delta\hat{H}=\Gamma\hat{a}_N^\dagger\hat{a}_1$ which shows a small deviation from the OBC configuration. In response to this, the unperturbed eigenstate $\left|\psi_{0}\right>_R$ will now change accordingly to $\left|\psi(\Gamma)\right>_R$ associated with a new eigenvalue that shifts from the zero point by $\Delta E$. In addition, to implement the perturbation $\Delta \hat{H}$ we use a third delay line which couples the first pulse to the last one in a non-reciprocal fashion. The strength of this coupling is then modulated accordingly to provide different values of the perturbation strength $\Gamma$. 

\begin{figure}
    \centering
    \includegraphics[width=0.7\textwidth]{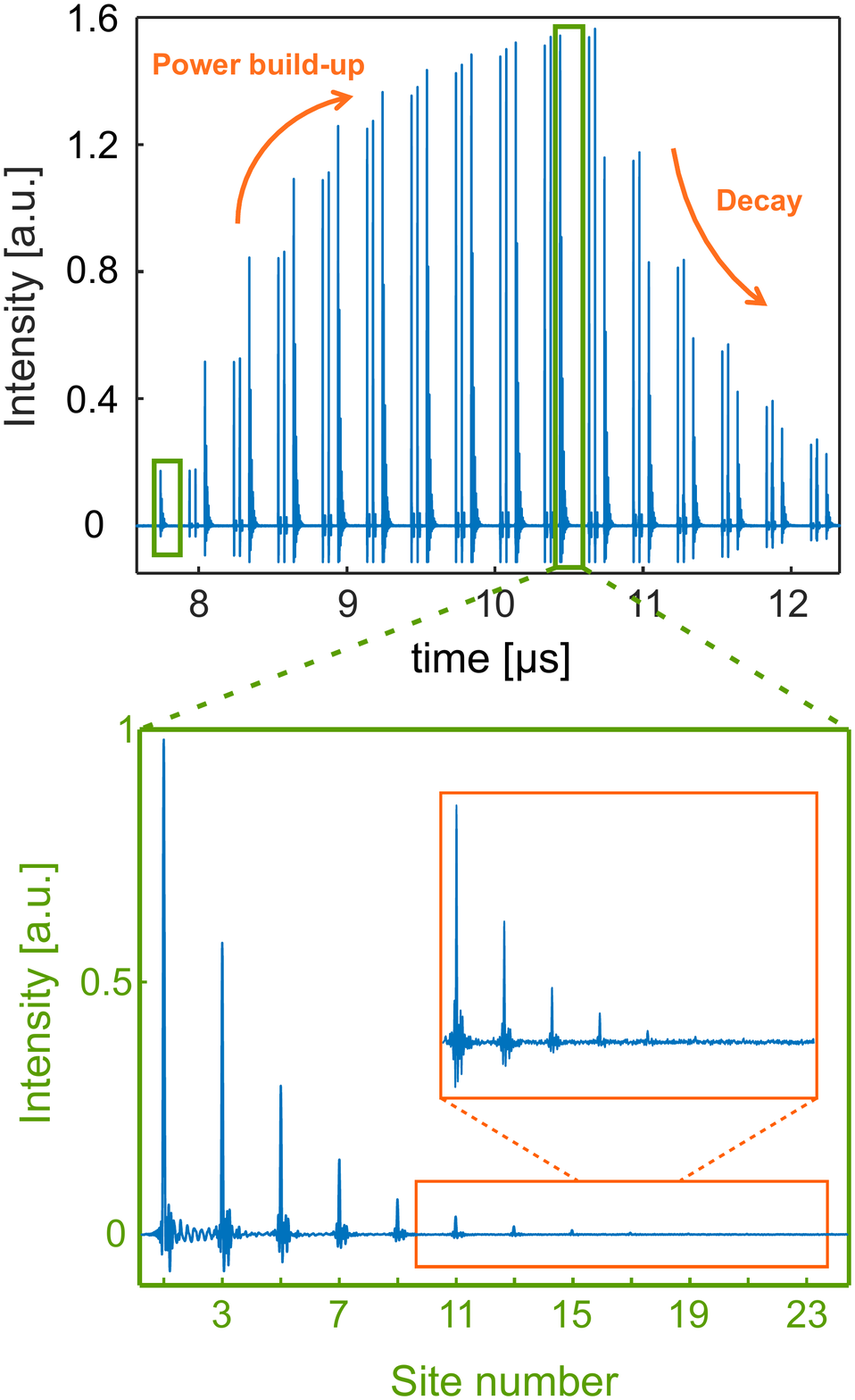}
    \caption{\textbf{Measurement procedure for the time-multiplexed NTOS.} Experimental time trace showing the pulse patterns at the output of the time-multiplexed resonator network for $N=23$. At the beginning ($t<10.5 \mu s$) optical pulses representing the zero eigenstate $\left|\psi_{0}\right>_R$ of the unperturbed Hamiltonian in Eq. \ref{eq:HN} (bottom green inset) are repeatedly injected into the closed cavity (power build-up regime). After this, the input path to the cavity is blocked while the delay lines are opened, allowing for the pulses to circulate inside the cavity and the delay lines. This results in a temporal decay of the input eigenstate for $t>10.5 \mu s$. By measuring these pulses and projecting them onto the left eigenstate of the unperturbed Hamiltonian $\left|\psi_{0}\right>_L$, we experimentally estimate the  shift in the zero eigenvalue $\Delta E$ associated with the Hatano-Nelson model resulting from the nonzero perturbation in the system.}
    \label{Pulses}
\end{figure}

In our experiments, we first initialize the system by shaping the amplitudes and phases of the input pulses to represent the zero eigenstate $\left|\psi_{0}\right>_R$ associated with the Hamiltonian in Eq. \ref{eq:HN}. Figure \ref{Pulses} shows an example of such pulses in the experiments concerning $N=23$ time-multiplexed resonators (the green inset depicts the zero eigenstate $\left|\psi_{0}\right>_R$). In order to increase the power of the pulses in the measurements, we repeatedly inject this pulse pattern into the closed cavity (with closed delay lines) which results in building up power inside the cavity (Fig. \ref{Pulses} for times $<10.5\mu s$). After this initialization, the input to the cavity is blocked so that the pulses start to circulate through the cavity and the delay lines according to the discrete-time evolution corresponding to the Hatano-Nelson model. Subsequently, at each time step that is defined by the integer multiples of the cavity round-trip time, we project the state of the network on the left eigenstate of the unperturbed Hamiltonian $\left|\psi_{0}\right>_L$. The perturbed eigenvalue can now be estimated from the decay rate of this projection per cavity round-trip. Using this, we then measure $\Delta E$ by calculating the difference between this new eigenvalue and the unperturbed one (see Methods). 

\begin{figure}
    \centering
    \includegraphics[width=1\textwidth]{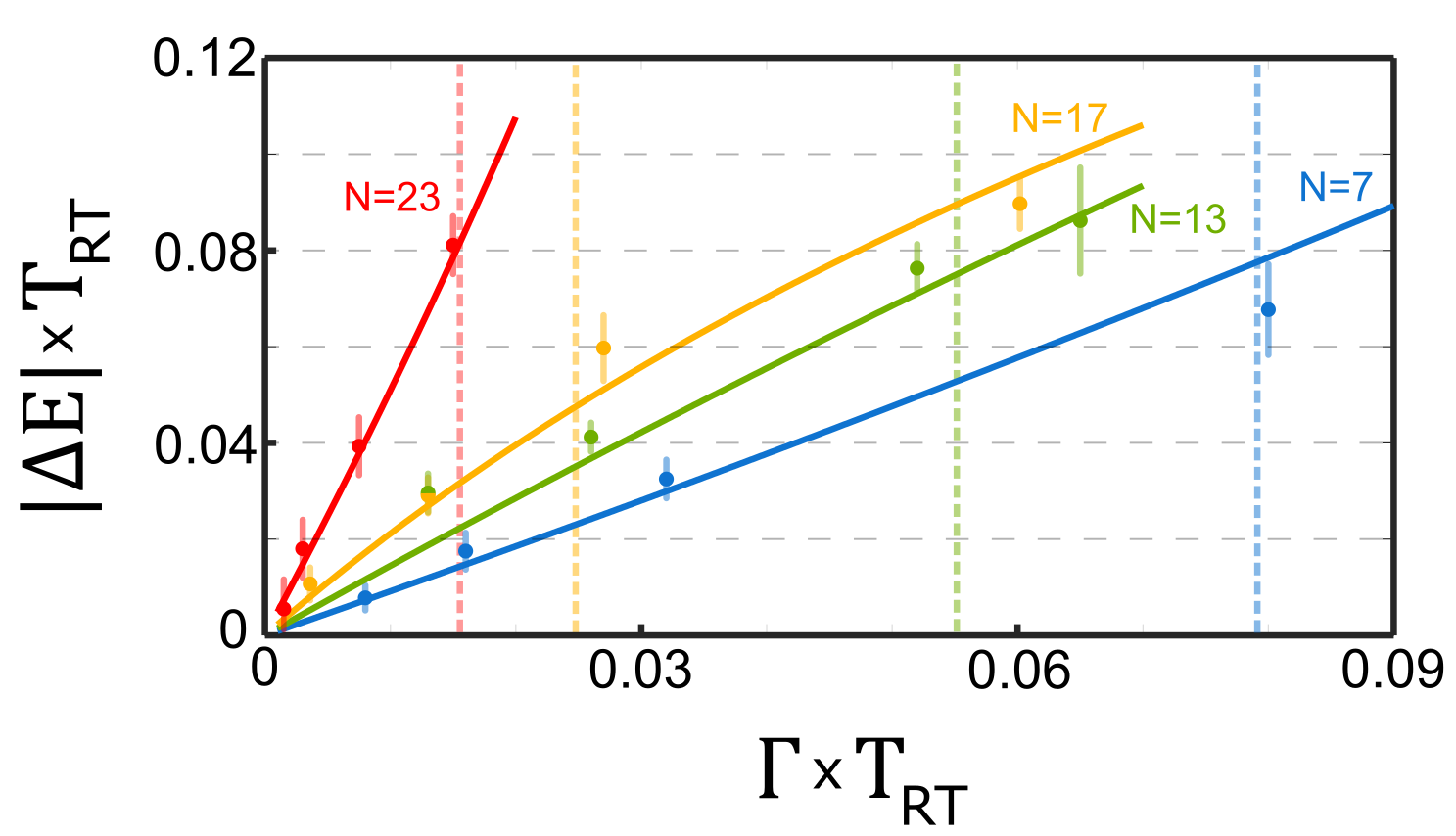}
    \caption{\textbf{Experimental demonstration of NTOS.} Experimentally measured shifts in the eigenvalue $\Delta E$ as the boundary coupling strength $\Gamma$ is perturbed from zero value (OBC conditions), for different lattice sizes $N=7, 13, 17$ and $23$. As evident in the figure, as long as $\Gamma$ is small enough, our NTOS responds linearly to the induced perturbations. However, as $\Gamma$ passes a threshold which depends on the size of the non-Hermitian topological lattice $N$, the change in the eigenvalue is no longer linear. The transition to this nonlinear regime is marked for each case in the figure by vertical dashed lines. Theoretically expected values are shown as solid curves. Here, $T_{RT}$ represents the round-trip time of the optical cavity.}
    \label{Experiment}
\end{figure}

Figure \ref{Experiment} displays experimentally measured values obtained from different lattices with various number of elements together with simulated results. For perturbations well below a critical value $\Gamma \ll \Gamma_C$, the NTOS exhibits a linear response with respect to the input parameter. However, for larger inputs, the perturbed eigenvalue associated with $\left|\psi(\Gamma)\right>_R$ is no longer real, signaling a crossover to the PBC where the sensor response is no longer linear \cite{kunst_biorthogonal_2018}. By increasing the perturbation further, the non-Hermitian skin effect breaks down and the eigenstates are no longer exponentially localized at the edge of the structure. Since the performance of the NTOS as a sensor is contingent upon this localization, it is crucial to avoid this non-Hermitian phase transition. Although in the thermodynamic limit $\Gamma_C$ tends to vanish, our analytical results show that for finite lattices its value remains nonzero and scales exponentially with $N$. In order to fully characterize our NTOS, we applied perturbations in a wide range of strengths spanning both below and above the aforementioned critical coupling. As shown in Fig. \ref{Experiment}, the experimentally measured results exhibit a linear system response to small $\Gamma$. For larger inputs, the sensor response eventually becomes nonlinear, hence setting the dynamic range of our demonstrated NTOS.

\begin{figure}
    \centering
    \includegraphics[width=1\textwidth]{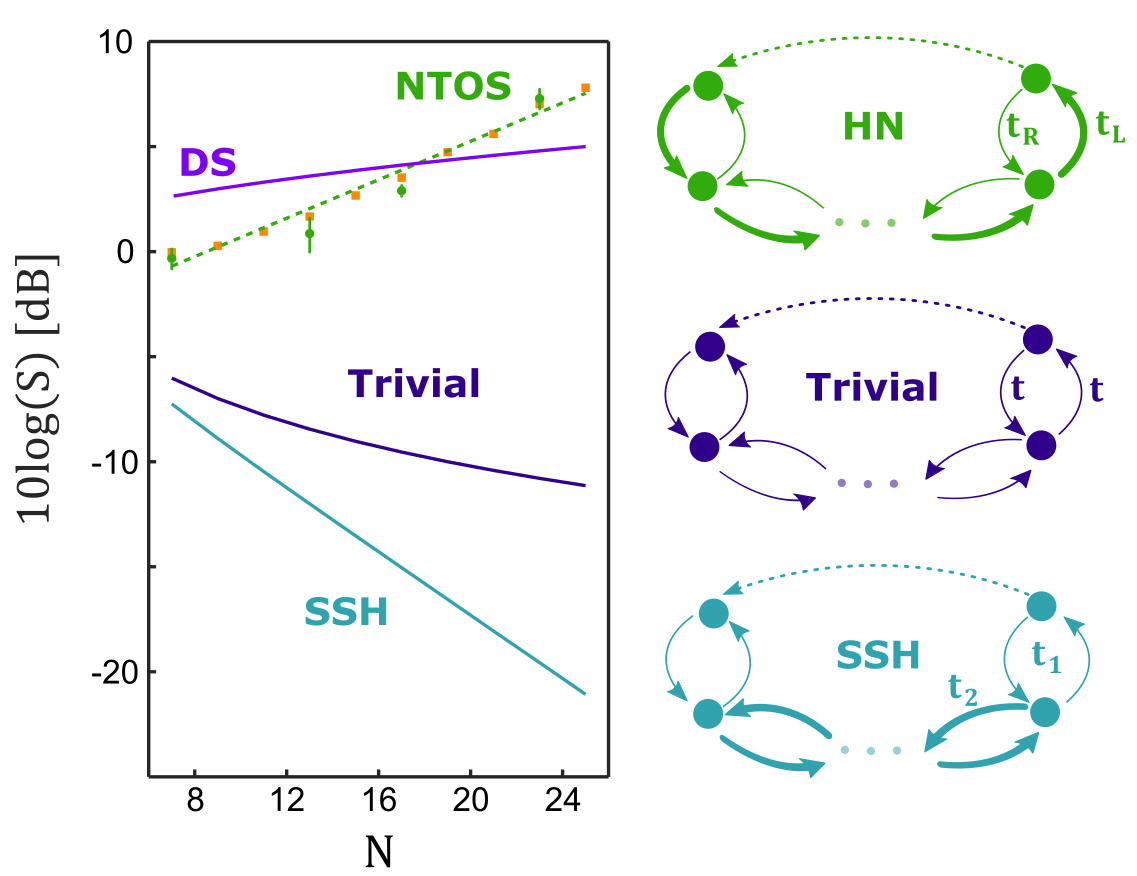}
    \caption{\textbf{Exponential enhancement in the sensitivity of the NTOS.} Experimentally obtained sensitivities $S$ of the NTOS for different lattice sizes $N$ are shown as green circles on the left plot. The corresponding theoretically predicted values are also depicted as orange squares. The data shows an exponential enhancement in the sensitivity $S$ as the NTOS lattice size grows (green dashed line). For comparison, we performed similar analysis for other types of lattices including a trivial lattice with uniform couplings as well as the Hermitian topological lattice represented by the SSH Hamiltonian (depicted on the right side of the figure). As shown in the plot, in sharp contrast to NTOS, such lattices tend to become less sensitive to their boundary conditions as the structure grows. The plot also displays the enhancement in the sensitivity resulting from distributing sensing (DS) techniques with $N$ sensing elements.}
    \label{Sens}
\end{figure}

To evaluate the performance of NTOS, we calculated the sensitivity defined as $S \equiv \partial E/\partial \Gamma$ using our measurement data in the small parameter regime $\Gamma \ll \Gamma_C$. Figure \ref{Sens} shows theoretically expected values along with experimental results for different lattice sizes $N$. From here, it is evident that the sensitivity of the NTOS grows exponentially with the size of this non-Hermitian topological system. Remarkably, the exponential enhancement of the sensitivity is known to arise in scenarios where the non-Hermitian topological winding number $\mathcal{W}$ defined as

\begin{equation}
    \mathcal{W} = \frac{1}{2\pi i} \int_{-\pi}^{\pi}dk \frac{\partial}{\partial k} \log\{\det[H(k)]\},
\label{eq:Winding}
\end{equation}

\noindent is nonzero \cite{budich_non-hermitian_2020}. Here, $H(k)$ denotes the Bloch Hamiltonian associated with the implemented lattice under PBC conditions. To corroborate this, we simulated the behavior of other types of lattices when subjected to the same perturbation $\Gamma$ in their boundary conditions as the NTOS studied here. We first consider the limiting case of the Hamiltonian in Eq. \ref{eq:HN} where the nearest-neighbor couplings become reciprocal $t_{\rm R}=t_{\rm L}$, resulting in a trivial system $\mathcal{W}=0$. As shown in Fig. \ref{Sens}, the sensitivity of a sensor implemented using a uniform lattice tends to deteriorate as $1/N$ with respect to the number of array elements. As a second example, we choose a Hermitian, but topologically non-trivial lattice, namely that associated with the Su-Schrieffer-Heeger (SSH) model \cite{su_solitons_1979}. When properly terminated, such a lattice also supports a pair of topological edge states that are localized in the open ends of the structure, in a way similar to the NTOS constructed in our experiments. However, unlike NTOS, the SSH Hamiltonian exhibits a trivial non-Hermitian winding number according to Eq. \ref{eq:Winding}. For this system, it can be shown that the sensitivity of the eigenvalues associated with such Hermitian edge states are in fact exponentially \textit{insensitive} to the changes in the boundaries of the array as $N$ grows (Fig. \ref{Sens}). These results hence confirm that the unusual enhancement in the sensing response observed in our experiments arises uniquely due to the synergy between non-Hermiticity and topology.

In summary, we have experimentally demonstrated enhanced sensitivity by non-Hermitian topological amplification based on the non-reciprocal Hatano-Nelson model. For various lattices with different number of elements, we characterized the response of the system as the shift in one of its eigenvalues as the boundary conditions change. While this response tends to saturate for perturbations larger than a critical limit, it tends to be linear for smaller range of values. The sensitivity parameter calculated using experimental data clearly exhibits an exponential growth with the lattice size $N$, in agreement with theoretical predictions. Using examples of other types of lattices, we showed that this peculiar enhancement arises due to the collaborative effect of non-Hermiticity and topology, something that does not occur for instance in Hermitian topological systems.

\section*{Methods}

\subsection*{Experimental Procedure}

To realize non-Hermitian topological sensors (NTOS), we construct the fiber-based time-multiplexed resonator network shown in Fig.~\ref{Schematic}. This network consists of a main cavity (yellow fiber) and three optical delay lines (blue fiber). We populate this network with optical pulses separated by a repetition period $T_{\rm R}\approx4 {\rm ns}$, and we choose the lengths of the delay lines to introduce couplings between these pulses. The $\pm1T_{\rm R}$ delay lines produce couplings between nearest-neighbor pulses in the cavity, while the $+(N-1)T_{\rm R}$ delay line, which is where we introduce perturbations, couples the first pulse in our synthetic lattice to the final pulse. While the main cavity can support up to $74$ pulses, we use the $+(N-1)T_{\rm R}$ delay line to set the size of the lattice under study, and we do not excite the unused time slots in the main cavity.

Prior to an experiment, we calibrate the electro-optic modulators (EOMs) in the network using the calibration procedure described in Supplementary Information Sec.~1b. We calibrate the EOM between the laser and the main cavity to carve the zero-mode of the unperturbed Hatano-Nelson lattice from the pulse train of the laser, while we calibrate the modulators in the $\pm1T_{\rm R}$ delay lines to implement the Hatano-Nelson model's asymmetric couplings. The $+(N-1)T_{\rm R}$ delay line also contains two EOMs (not shown in Fig.~\ref{Schematic}), which control the strength of the perturbation between the first and final sites of the HN lattice. We calibrate the throughput of these modulators to set the perturbation strength for any given experiment.

After completing our calibration, we begin an experiment by injecting the Hatano-Nelson zero-mode into the network for 10 roundtrips, which allows the power in the zero-mode to resonantly build up within the cavity. During this time, we leave the IMs in the $\pm1T_{\rm R}$ delay lines biased to minimum throughput so that we do not couple neighboring pulses through these delay lines. After 10 roundtrips, we stop injecting the zero-mode and we turn on the couplings in the $\pm1T_{\rm R}$ delay lines. We save a trace of the cavity ring-down, and we repeat this measurement on the order of 50 times to generate statistics for our data analysis.

In addition to injecting the zero-mode into our network, we also inject a single pulse into one of the unused time slots of the main cavity. We leave this single pulse uncoupled to the surrounding time slots so that this pulse decays at the intrinsic decay rate of just the main cavity. In the absence of the perturbation, this is the same decay rate that we would expect for the zero-mode of the Hatano-Nelson model. Therefore, this auxiliary pulse acts as a reference from which we can extract the change in the decay rate of the zero-mode due to the perturbation.

\section*{Acknowledgments}
The authors acknowledge support from ARO Grant W911NF-23-1-0048 and NSF Grants No. 1846273 and 1918549. The authors wish to thank NTT Research for their financial and technical support.

\section*{Author Contributions}

All authors contributed to the writing of this manuscript.

\section*{Competing Interests}

The authors declare no competing interests with regards to the publication of this work.

\section*{Data Availability}

The data used to generate the plots and results in this paper is available from the corresponding author upon reasonable request.

\section*{Code Availability}

The code used to analyze the data and generate the plots for this paper is available from the corresponding author upon reasonable request.

\printbibliography

\end{document}